\documentclass{article}
\usepackage{amsmath, amssymb, amsthm}
\usepackage{hyperref}
\usepackage{xcolor}

\hypersetup{
    colorlinks = true,
    linkcolor = blue,
    urlcolor  = blue,
    citecolor = blue,
    anchorcolor = blue
}

\begin{document}

\title{Dynamic Many-Valued Logic Systems in Theoretical Economics}
\renewcommand{\thefootnote}{\ifcase\value{footnote}\or\textdagger\fi}
\author{Daniel Lu\footnote{\href{https://orcid.org/0009-0003-7803-1683}{ORCID: 0009-0003-7803-1683}; \href{mailto:Daniel.Lu@warwick.ac.uk}{Daniel.Lu@warwick.ac.uk}; \textit{Warwick Business School}, University of Warwick}}
\date{March, 2024}
\maketitle
\begin{abstract}
    This paper is an original attempt to understand the foundations of economic reasoning. It endeavors to rigorously define the relationship between subjective interpretations and objective valuations of such interpretations in the context of theoretical economics. This analysis is substantially expanded through a dynamic approach, where the truth of a valuation results in an updated interpretation or changes in the agent's subjective belief regarding the effectiveness of the selected action as well as the objective reality of the effectiveness of all other possible actions (i.e. consequence realization). Complications arise when the economic agent is presented with a set of actions that render ambiguous preference, or when the effectiveness of an action cannot be perceived upon its selection, thereby necessitating a different theory of choice and consequence realization. 
\end{abstract}
\section{Introduction}
Suppose you are interested in selecting a portfolio that maximizes future returns. We may express each portfolio as an element, $p_{i}$, in the set $\mathbb{P}$. As opposed to traditional statistical approaches, we shall view each $p_{i}$ as a assertion. For example, $p_{0}$ can be defined as saying “Investing 10\% of one's wealth in asset A, 20\% in asset B... would result in the maximization of future returns." By defining each member of $\mathbb{P}$ in this manner, we may assign truth values to each sentence. According to the law of the excluded middle, either a statement is true or false; either it is the case that a portfolio results in the maximization of future returns, or it does not. In this naive model, the investor is confronted with a simplistic computational task where a true $p_{i}$ is selected. To express the selection of a portfolio, we may use the sentence $s$, which is syntactically equivalent to saying “This portfolio shall be selected." To formally denote a valid selection, we shall write $p_{i}\rightarrow s$. Under classical logic, the meaning of the statement $p_{i}\rightarrow s$ has no bearing on its logical form. In this case, the assertion in question is an instantiation of an “If...then" relationship. We could equally replace $p_{i}$ and $s$ with arbitrary variables such as $x$ and $y$ without necessarily dispensing with the fundamental logical abstraction. In fact, classical logic posits that the truthfulness of the statements $p_{i}$ and $s$ mirror that of the cohesive statement $p_{i}\rightarrow s$ by following the law of the excluded middle. This invites us to examine two epistemological question: can a portfolio's ability to maximize future returns be $true$ in the same way that its selection is $true$? (Or, in the same way that the statement $p_{i}\rightarrow s$ is interpreted as being a $true$ assertion?) Secondly, can the logical form of a statement be entirely separate its meaning? 
\section{Motivation}
This paper is predicated on the position that the answers to the preceding questions are no. Classical logic offers an inadequate explanation of how objects can possess truth values and be grouped through logical connectives. This is because reasoning within an object language is intrinsically different from reasoning about its semantic properties, and the semantic properties underpinning the ascription of such properties. The primary motivation behind the creation of many-valued logical systems, where one abandons the absoluteness of binary notions such as $True$ or $False$, is to elucidate this difference that underpins the essence of a logical system.

An analogy for this is to imagine a box in which a person is asked to examine the semantic properties of a list of statements on a piece of paper, whatever they may be. Another person outside the box, who observes this, can be asked about the semantic properties of the reasoning used by the person within the box. The nature of these semantic properties, whether true, false, unknown, etc. and their interrelations can be flexibly designed in a logical manner that fits desired purposes.

Returning to the introductory example, let us consider two portfolios, $p_{k}$ and $p_{j}$. By interpreting both assertions individually, we may examine their truthfulness in terms of whether they are $conducive$ to the goal of maximizing future returns. Whereas the formation of a complex sentence, such as $p_{k}$ $\lor$ $p_{j}$, should be deemed as an entirely new statement that can be interpreted as asking whether either portfolio would be $selected$. In other words, an interpretation on such a sentence occupies an entirely different epistemological realm, and their respective $truths$ are not structurally identical. Any truth value attributable to the statement  $p_{k}$ $\lor$ $p_{j}$ is an assertion regarding the selection of either portfolio, whereas any truth value attributable to each individual portfolio is an assertion pertaining to the validity of its maximal quality. 

Additionally, instances where the logical form of a sentence cannot adequately capture its inherent meaning are prevalent. Most notably, propositional logic (PL, a formal language that does not feature quantifiers, such as $\forall$ or $\exists$) cannot be used to express the validity of syllogisms such as: \\

$A$:= All investment options with a Sharpe ratio greater than 1 are considered to have a good risk-adjusted return. 

$B$:= Mutual Fund $\mathfrak{K}$ has a Sharpe ratio of 1.5. 

$C$:= Therefore, Mutual Fund $\mathfrak{K}$ is considered to have a good risk-adjusted return.
\\
This is because PL does not have the expressive power of quantifiers and therefore does not allow the derivation of any statement \(C\) from the validity of statements \(A\) and \(B\). 

Another example would be the underlying epistemology of the conditional “If...then" relation. In classical logic, a conditional relation is a false assertion if and only if a false consequent arises from a true antecedent. However, there exist conditionals where the truth of the antecedent is necessary though insufficient to guarantee the truth of the consequent. For example, consider the statement  “If you invest in stock $X$ then you will reap its returns."  Under the usual interpretation of the “If...then" relation, if an investment is made in stock $X$ and future losses are permanently incurred, then the conditional relation is false. However, the relation only suggests that an investment is necessary though insufficient for one to reap its returns. In other words, the conditional relation is false if and only if the antecedent is false and the consequent is true, since the consequent is necessarily contingent upon the truth of the antecedent. 

First order logic lacks the expressive power of modal logic, which features nuanced notions such as \( \textit{necessarily true} \) or \( \textit{possibly true} \). This is symptomatic of an inherent problem in the pathological insistence on the notion that meaning is distinct from logical form when, in fact, they are inextricably linked. As opposed to making further abstractions regarding the logical form of sentences through the introduction of additional logical operators, (e.g., going beyond quantifiers such as \( \forall \) or \( \exists \) or modal operators like \( \Box p \) a.k.a it is necessarily the case that \(p\)), this paper assumes that each sentence, whether complex or atomic, could assume any truth value as they may address fundamentally different epistemic questions.

The objective of this essay is to approach the application of dynamic many-valued logic in theoretical economics in an abstract, rigorous, and accessible manner. In addition, this essay endeavors to offer a fundamental understanding of economic reasoning. We shall begin by introducing the basic definitions that make up a logical system.\\

\noindent\textbf{Definition 2.1} An atomic sentence, \(\varphi\), is a letter for which a semantic truth value can be assigned. We denote the set of all atomic sentences with \( \mathcal{A} \). \\

\textbf{Recall} from the introductory example the set $\mathbb{P}$ which stands for the collection of all assertions regarding the optimal portfolio. \\

\noindent\textbf{Definition 2.2} A semantic truth value, \( T_k \), can be construed as an arbitrary denotation that says something about the properties of an object. Let there be a set such that \( T_k \) belongs to the collection \(\mathbb{T}_{d}\), or equivalently, \( \{ T_i \}_{i=0}^{j} \) for some \( j\) or \(d\in \mathbb{N} \), where each \( T_i \) is a truth value. Formally, we define \( T_k \) as follows:
\[
T_k \in \{ T_i \ | \ i \in \{0, \ldots, j\}, j \in \mathbb{N} \}
\]

\textbf{Recall} from the introductory example how a truth value can be used to denote whether an individual portfolio is capable of rendering the maximum level of future returns, or indicate whether it has been selected.\\

To interpret a sentence is to acknowledge an unambiguous many-to-one mapping from sentence letters to a unique truth value. Though multiple sentences may be understood as having the same truth value, it is assumed that the same sentence cannot be interpreted in a singular way that results in the emergence of multiple truth values. \\

\noindent\textbf{Definition 2.3} An interpretation is a function that assigns an atomic sentence a unique truth value in $\mathbb{T}_{d}$ for some $d$ in $\mathbb{N}$:
\[ I: \mathcal{A} \rightarrow \mathbb{T}_{d}\]

\textbf{Example: } $I$$(p_{k})=T$ i.e. Portfolio k is maximal. \\

\noindent\textbf{Definition 2.4} 
A complex sentence consists of atomic sentences that are grouped by logical connectives. We denote the set of all complex sentences with \( \mathcal{S} \).\\

\textbf{Example: }($p_{k}$ $\lor$ $p_{j}$) $\in$ $\mathcal{S}$ \\

\noindent\textbf{Definition 2.5} (Logical Connective) A logical connective ($\oplus_{j}$) is a function that specifies the logical relationship between a sequence of atomic sentence letters of arity $k$ in $\mathbb{N}$ and their respective roles in a complex sentence. For some $j$ in $\mathbb{N}$, there exists a $\oplus_{j}$ such that:
\[
\oplus_{j} : \mathcal{A}^k \rightarrow \mathcal{S}
\]

\textbf{Example: $\lor$}($p_{k}$, $p_{j}$) := ($p_{k}$ $\lor$ $p_{j}$) \\

\noindent\textbf{Definition 2.6} A valuation function, $V_{I}$, is a function that assigns a semantic truth value in $\mathbb{T}_{d}$ for some $d$ in $\mathbb{N}$ to a complex sentence on the basis of the truth values of its individual parts, regardless of whether or not the range of the complex sentence's truth value coincides with that of its atomic counterparts. 
\[
V_{I}: \mathcal{S} \rightarrow \mathbb{T}_{d}
\]

\textbf{Example: }$V_{I}$($p_{k}$ $\lor$ $p_{j})=T$ i.e. The selection of portfolios $k$ or $j$ is true.\\

Although classical logic conflates the codomain of the valuation function with the codomain of the interpretational function, it is necessary to specify an important metalogical distinction between the two. This distinction is necessary because logical connectives may be used to express arbitrary semantic consequences given an arbitrary relationship between the truth values of multiple atomic sentences. Let $\psi$ and $\varphi$ be sentence letters and $\oplus_0$ an arbitrary binary connective; let the codomain of an interpretational function be $\mathfrak{\{M,F\}}$; let that of the connective be $\{A,B,C,D\}$. This allows us to define $\oplus_0$ in the following arbitrary manner:

\[
\varphi \oplus_{0}\psi
\]
\begin{center}
\begin{tabular}{|c|c|c|}
\hline
\(\varphi \smallsetminus \psi\) & \(I(\psi) = \mathfrak{M}\) & \(I(\psi) = \mathfrak{F}\) \\
\hline
\(I(\varphi) = \mathfrak{M}\) & \(V_{I}\left( \varphi \oplus_{0}\psi \right)= A\) & \(V_{I}\left( \varphi \oplus_{0}\psi \right)= C\) \\
\hline
\(I(\varphi) = \mathfrak{F}\) & \(V_{I}\left( \varphi \oplus_{0}\psi \right)= B\) & \(V_{I}\left( \varphi \oplus_{0}\psi \right)= D\) \\
\hline
\end{tabular}
\end{center}
\[\psi \oplus_{0}\varphi\]
\begin{center}
\begin{tabular}{|c|c|c|}
\hline
\(\psi \smallsetminus \varphi\) & \(I(\varphi) = \mathfrak{M}\) & \(I(\varphi) = \mathfrak{F}\) \\
\hline
\(I(\psi) = \mathfrak{M}\) & \(V_{I}\left( \psi \oplus_{0}\varphi \right)= A\) & \(V_{I}\left( \psi \oplus_{0}\varphi \right)= C\) \\
\hline
\(I(\psi) = \mathfrak{F}\) & \(V_{I}\left( \psi \oplus_{0}\varphi \right)= B\) & \(V_{I}\left( \psi \oplus_{0}\varphi \right)= D\) \\
\hline
\end{tabular}
\end{center}
\[\varphi \oplus_{0}\varphi\]
\begin{center}
\begin{tabular}{|c|c|}
\hline
\(\varphi\) & \(\varphi \oplus_{0}\varphi\) \\
\hline
\(I(\varphi) = \mathfrak{M}\) & \(V_{I}\left( \varphi \oplus_{0}\varphi \right)= A\) \\
\hline
\(I(\varphi) = \mathfrak{F}\) & \(V_{I}\left( \varphi \oplus_{0}\varphi \right)= D\) \\
\hline
\end{tabular}
\end{center}
\[\psi \oplus_{0}\psi\]
\begin{center}
\begin{tabular}{|c|c|}
\hline
\(\psi\) & \(\psi \oplus_{0}\psi\) \\
\hline
\(I(\psi) = \mathfrak{M}\) & \(V_{I}\left( \psi \oplus_{0}\psi \right)= A\) \\
\hline
\(I(\psi) = \mathfrak{F}\) & \(V_{I}\left( \psi \oplus_{0}\psi \right)= D\) \\
\hline
\end{tabular}
\end{center}
The preceding tables are known as truth-table. A truth table consists of different interpretations of two atomic sentences on the rows and columns. Each square in the middle of the table shows the outcome of a valuation on a binary connective when the interpretation of each sentence is taken into consideration (e.g. $V_{I}\left( \psi \oplus_{0}\psi \right)= A$). The exact meaning of each truth value in $\mathfrak{\{M,F\}}$ or $\{A,B,C,D\}$ is irrelevant here, as they may be used to address different epistemic questions. 

Note that the limit to the number of semantic truth values of the valuation of a $j$-nary connective, where each object can have $n$ semantic interpretations, is $n^j$. When this is the case, it is possible to derive the exact interpretation of every sentence letter solely on the basis of the truth value of the valuation. For example, the valuation $V_I (\psi \oplus_0 \varphi) = C$ allows us to derive that $I(\psi) = \mathfrak{M}$ whereas $I(\varphi) = \mathfrak{F}$.
\\

\section{Note to Reader}
The reader should bear the following in mind. Firstly, a formal language cannot be understood intuitively (Falkenstein et al.). When applied to theoretical economics, a formal language can be thought of as a counter-intuitive rendition of the nuances of economic reasoning. When an $informal$ justification is provided, the reader should bear in mind that the reasoning or idiosyncratic notations being used should be interpreted as aids to one's comprehension, but not an accurate representation of what is being conveyed. The reader should consider this work as an exercise in philosophy. It is an attempt to understand the interplay between economic reasoning and the results of its application. Due to the density of the notations and concepts featured, the reader may be frequently asked to recall concepts in preceding sections to aid one's understanding.

\section{Six-Valued Logic System (SVL)}

Consider the following set of truth values which specifies the codomain the interpretational function: 
\[ \mathbb{T}_{0} =\{T,F,U\} \]

Where \( T \) stands for “Interpretationally True”, \( F \) stands for “Interpretationally False”, and \( U \) stands for “Interpretationally Unknown”. These are truth values that can only be ascribed to atomic setences. In contrast, consider the codomain for the valuation function: 
\[ \mathbb{T}_{1}=\{t,f,u\} \]
Where \( t \) stands for “Cohesively True”, \( f \) stands for “Cohesively False”, and \( u \) stands for “Cohesively Unknown”. The exact meaning of each truth value is irrelevant here as we are trying to incorporate the epistemology of each statement into a particular logical interpretation. It is only necessary for us to understand that $\mathbb{T}_{1}$ stands for the collection of all truth values that can be assumed by complex sentences. 

Consider the following logical connectives in this propositional calculus: namely, not (\( \neg_{SVL} \)), and (\( \land_{SVL} \)), or (\( \lor_{SVL} \)), and if…then (\( \rightarrow_{SVL} \)). To specify what the logical connectives say about the relationship between sentence letters and the truth value/valuation of the whole sentence, the following tables can be created.
\begin{center}
\[
\neg_{SVL}\varphi
\]
\[
\begin{array}{|c|c|}
\hline
 \varphi &  \neg_{SVL} \varphi \\
\hline
I(\varphi)=T & V_I (\neg_{SVL} \varphi)=f \\
\hline
I(\varphi)=F & V_I (\neg_{SVL} \varphi)=t \\
\hline
I(\varphi)=U & V_I (\neg_{SVL} \varphi)=u \\
\hline
\end{array}
\]
\end{center}
\begin{center}
\[
\varphi \land_{SVL} \psi
\]
\[
\begin{array}{|c|c|c|c|}
\hline
\varphi \smallsetminus \psi & I(\psi)=T & I(\psi)=F & I(\psi)=U \\
\hline
I(\varphi)=T & V_I (\varphi \land_{SVL} \psi)=t & V_I (\varphi \land_{SVL} \psi)=f & V_I (\varphi \land_{SVL} \psi)=u \\
\hline
I(\varphi)=F & V_I (\varphi \land_{SVL} \psi)=f & V_I (\varphi \land_{SVL} \psi)=f & V_I (\varphi \land_{SVL} \psi)=f \\
\hline
I(\varphi)=U & V_I (\varphi \land_{SVL} \psi)=u & V_I (\varphi \land_{SVL} \psi)=f & V_I (\varphi \land_{SVL} \psi)=u \\
\hline
\end{array}
\]
\end{center}

\begin{center}
\[
\varphi \lor_{SVL} \psi
\]
\[
\begin{array}{|c|c|c|c|}
\hline
\varphi \smallsetminus \psi & I(\psi)=T & I(\psi)=F & I(\psi)=U \\
\hline
I(\varphi)=T & V_I (\varphi \lor_{SVL} \psi)=t & V_I (\varphi \lor_{SVL} \psi)=t & V_I (\varphi \lor_{SVL} \psi)=t \\
\hline
I(\varphi)=F & V_I (\varphi \lor_{SVL} \psi)=t & V_I (\varphi \lor_{SVL} \psi)=f & V_I (\varphi \lor_{SVL} \psi)=u \\
\hline
I(\varphi)=U & V_I (\varphi \lor_{SVL} \psi)=t & V_I (\varphi \lor_{SVL} \psi)=u & V_I (\varphi \lor_{SVL} \psi)=u \\
\hline
\end{array}
\]
\end{center}

\begin{center}
\[
\varphi \rightarrow_{SVL} \psi
\]
\[
\begin{array}{|c|c|c|c|}
\hline
\varphi \smallsetminus \psi & I(\psi)=T & I(\psi)=F & I(\psi)=U \\
\hline
I(\varphi)=T & V_I (\varphi \rightarrow_{SVL} \psi)=t & V_I (\varphi \rightarrow_{SVL} \psi)=f & V_I (\varphi \rightarrow_{SVL} \psi)=f \\
\hline
I(\varphi)=F & V_I (\varphi \rightarrow_{SVL} \psi)=u & V_I (\varphi \rightarrow_{SVL} \psi)=u & V_I (\varphi \rightarrow_{SVL} \psi)=t \\
\hline
I(\varphi)=U & V_I (\varphi \rightarrow_{SVL} \psi)=u & V_I (\varphi \rightarrow_{SVL} \psi)=u & V_I (\varphi \rightarrow_{SVL} \psi)=t \\
\hline
\end{array}
\]
\end{center}

An $informal$ justification for each arbitrary table can be offered:
\begin{itemize}
    \item The Negation Table specifies that if a sentence is known to be true, then its negation must be false, and vice-versa. However, if the sentence’s truth value cannot be determined, then neither can its negation. More specifically, when the interpretation of $\varphi$ is $U$, the valuation of $(\varphi \lor \neg_{SVL}\varphi)$ is $u$. This may seem counter-intuitive at first; one may argue that although $\varphi$ may be unknown, either $\varphi$ or $\neg_{SVL}\varphi$ is true. However, that is the exact subject in question; we cannot determine whether $\varphi$ is true or not true. Hence, the valuation must be $u$.
    
    \item The And Table states that if one of the interpretations of the sentences is false, then the valuation of the whole sentence must also be cohesively false. However, if one is true and the other is unknown, then the valuation is also unknowable since the interpretation of a sentence with an unknowable valuation can either be true or false. To note the consistency of this multi-valued system, one may argue that $(\varphi \land_{SVL} \neg\varphi)$ must be false even if the interpretation of $\varphi$ is unknown because $\varphi$ cannot have two values at once. However, if $(\varphi \land_{SVL} \neg_{SVL}\varphi)$ were false, then either $\varphi$ is false, or its negation is false. But since the valuation of $\varphi$ is unknown, then the valuation of its negation must be unknown as well, hence the valuation of $(\varphi \land_{SVL} \neg_{SVL}\varphi)$ is not false, but rather, unknown. $U/u$ must be interpreted as truth values whose definitions are independent of the meaning of other truth values; absolute ambiguity is a valid logical state.
    
    \item The Or Table states that if one of the sentence values is interpreted as being true, then the whole sentence is true, regardless of whether the valuation of the other sentence is true, false, or unknowable. However, if both are false then the whole sentence is false since neither sentence is true. But if one sentence is false/unknowable, and the interpretation of the other is unknowable, then the valuation of the whole sentence is also unknown.
    
    \item The If…Then Table states that if $\varphi$ is interpreted to be true, then $\psi$ must also be true: in which case $\psi$ cannot be false or unknown, otherwise that would result in the whole sentence being valued as false. However, if $\varphi$ were false or unknown, then what can be reasonably inferred is that the valuation for the whole construction is also unknown. This is because the If…Then statement only specifies the resulting value of $\psi$ when and only when $\varphi$ is true. Otherwise, it may either be true or false, and only its unknowability can be affirmed.
\end{itemize}

We may now proceed with this analysis by developing a set of inference rules through a reversal of the direction in which we interpret the previous tables. This allows us to examine what can be said of the truth values of an individual sentence(s) given the valuation of the complex sentence, and in some instances, given, also, the interpretation of a different sentence. Let $\Gamma$ represent a set of premises from which we infer a conclusion, $\phi$. An inference is valid insofar as all the premises in $\Gamma$ are true and the resulting conclusion is not false. A premise must be explicitly specified to have meaning. For example, suppose we are privy to the fact that the valuation of a particular interpretation of the sentence “$\neg\varphi$” is false. From this premise, we derive that that exact interpretation must have assessed “$\varphi$” to be true. This practice of explicitly enunciating the truth values of premise sentences allows for greater flexibility and coherence in the analysis of many-valued logic.

We shall define the premises in a way that ascertains the individual interpretations of the atomic sentences and infers what the valuation on the interpretation says about a particular logical connective. This requires us to define an Equivalence Valuation Function, \(\mathfrak{v} \), which ascertains whether the individual interpretations or valuations are semantically equivalent to their purported truth values. The codomain of this function is \(\mathfrak{\{T,F\}} \). 
For example: 

\begin{center}
From \( \Gamma := \{ \mathfrak{v} (I(\varphi),T)=\mathfrak{T},\mathfrak{v}(I(\psi),T)=\mathfrak{T} \} \) infer \( \mathfrak{v}(V_I (\varphi \land_{SVL} \psi),t) = \mathfrak{T}\) \\

\end{center}
\noindent\textbf{Definition 4.1 (Equivalence Valuation Function)}: Consider the logical system defined by the atomic sentences ($\mathcal{A}$), complex sentences ($\mathcal{S}$), the interpretation function $I$, the valuation function $V_I$, and the set of semantic truth values ($\mathbb{T}_{d}$) as outlined in Definitions 2.1 to 2.6. The equivalence valuation function, $\mathfrak{v}$, is a mechanism for assessing the fidelity of the semantic assignments made by $I$ and $V_I$ against a predetermined equivalence criterion with respect to the truth value $T_k$ in $\mathbb{T}_{d}$. Formally, $\mathfrak{v}$ operates as follows:

\begin{itemize}
    \item For any atomic sentence $\varphi \in \mathcal{A}$, and any truth value $T_k$, $\mathfrak{v}$ evaluates the interpretation $I(\varphi)$ against $T_k$ based on a predefined equivalence relation. Specifically, $\mathfrak{v}(I(\varphi), T_k)$ yields $\mathfrak{T}$ if the truth value assigned by $I$ to $\varphi$ is deemed equivalent to $T_k$, and $\mathfrak{F}$ otherwise.
    \item For any complex sentence $\psi \in \mathcal{S}$, and any truth value $T_k$, $\mathfrak{v}$ assesses the valuation $V_I(\psi)$ against $T_k$ according to the same equivalence relation. In this case, $\mathfrak{v}(V_{I}(\psi), T_k)$ results in $\mathfrak{T}$ if the truth value determined by $V_I$ for $\psi$ is considered equivalent to $T_k$, and $\mathfrak{F}$ otherwise.
\end{itemize}

Thus, the function $\mathfrak{v}$ can be expressed in terms of its domain and range as follows:
\[
\mathfrak{v} : (\mathcal{A} \cup \mathcal{S}) \times \mathbb{T}_{d} \rightarrow \{\mathfrak{T,F}\}
\]
where the domain consists of pairs of either atomic or complex sentences and a particular semantic truth value $T_k$, and the range is the set $\{\mathfrak{T,F}\}$, representing the truthfulness of the equivalence relation between the sentence's assigned truth value and $T_k$.

Notice that when the Equivalence Valuation Function is used, a fraktur capital \( T \), \( \mathfrak{T} \), is used to denote the validity of the statement “\( I(\psi)=T \)” in lieu of \( T \) or \( t \). This reflects a subtle though important distinction between statements within and about a metalanguage (i.e. the meta-metalanguage). Even if a many-valued logical system features more than two truth values, the concept of an equivalence relation is fundamentally binary. For example, it is either the case that the interpretation of \( \psi \) assumes a particular truth value, whether it may be \( T,F \) or \( U \), or it is not.
The previous 6-valued logic system offers a variety of inference relations. An incomplete list of which is presented here. \\

Negation Inferences:
\begin{center}

From \( \Gamma := \{ \mathfrak{v}(V_I (\neg_{SVL} \varphi),f)=\mathfrak{T} \} \) infer \(  \mathfrak{v}(I(\varphi),T)=\mathfrak{T} \) \\

From \( \Gamma := \{ \mathfrak{v}(V_I (\neg_{SVL} \varphi),t)=\mathfrak{T} \} \) infer \(  \mathfrak{v}(I(\varphi),F)=\mathfrak{T} \) \\

From \( \Gamma := \{ \mathfrak{v}(V_I (\neg_{SVL} \varphi),u)=\mathfrak{T}\} \) infer \(  \mathfrak{v}(I(\varphi),U)=\mathfrak{T} \) \\
\end{center}

And Inferences:

\begin{center}

From \( \Gamma := \{ \mathfrak{v}(V_I (\varphi \land_{SVL} \psi),t)=\mathfrak{T},\mathfrak{v}(I(\psi),T)=\mathfrak{T}\} \) infer \(  \mathfrak{v}(I(\varphi),T)=\mathfrak{T} \) \\

From \( \Gamma := \{ \mathfrak{v}(V_I (\varphi \land_{SVL} \psi),t)=\mathfrak{T},\mathfrak{v}(I(\psi),T)=\mathfrak{T} \} \) infer \(  \mathfrak{v}(I(\varphi),T)=\mathfrak{T} \) \\

From \( \Gamma := \{ \mathfrak{v}(V_I (\varphi \land_{SVL} \psi),f)=\mathfrak{T},\mathfrak{v}(I(\psi),T)=\mathfrak{T} \} \) infer \(  \mathfrak{v}(I(\varphi),F)=\mathfrak{T} \) \\

From \( \Gamma := \{ \mathfrak{v}(V_I (\varphi \land_{SVL} \psi),f)=\mathfrak{T},\mathfrak{v}(I(\varphi),T)=\mathfrak{T} \} \) infer \(  \mathfrak{v}(I(\psi),F)=\mathfrak{T} \) \\

\(\vdots\)
\end{center}

\noindent\textbf{Definition 4.2 (Entailment):} Entailment is a relationship between a set of statements or propositions (premises) and an individual statement (conclusion), where the validity of the premises logically guarantees the validity of the conclusion. This concept is crucial for understanding logical inference and reasoning processes.

To denote a valid inference, where a set of premises \( \Gamma \) logically leads to a conclusion \( \phi \), we write \( \Gamma \models \phi \). This notation signifies that if all the statements in \( \Gamma \) are true, then \( \phi \) must also be true. The symbol \( \models \) represents the entailment or logical consequence, indicating that the conclusion \( \phi \) is an entailment of the premises in \( \Gamma \).

\subsection{Note on $\mathfrak{v}$ and the Law of the Excluded Middle}

The law of the excluded middle states that for all statements, either it is the case that the statement is true, or it is not the case that the statement is true, or it is the case that a statement is false, or it is not the case that a statement is false. To say that it is not the case that a statement is true is equivalent to saying that it is the case that a statement is false, and to say that it is not the case that a statement is false is equivalent to saying that it is the case that a statement is true.

To express this definition formally, let $\phi$ be an arbitrary sentence letter. Let $I(\phi)$ be an interpretational function that assigns a particular truth value to the sentence letter. Let the domain of $I$ be any sentence letter and let its codomain be $\{T, F\}$. Recall the definition of an equivalence valuation function whose domain is any equivalence relation between an interpretation and a semantic property and whose codomain is strictly limited to $\{\mathfrak{T}, \mathfrak{F}\}$. We may therefore formalize the law of the excluded middle in the following manner.

\begin{equation} \label{lem1}
\forall \phi (((\mathfrak{v}(I(\phi), T) = \mathfrak{T} ) \lor (\mathfrak{v} (I(\phi), F) = \mathfrak{T} )) \lor ( \mathfrak{v}(I(\phi), T) = \mathfrak{F} )) \lor ( \mathfrak{v}(I(\phi) ,F) = \mathfrak{F} ))
\end{equation}
\begin{equation} \label{lem2}
 (\mathfrak{v}\left(I(\phi), T\right) = \mathfrak{F})  \Leftrightarrow (\mathfrak{v}\left(I(\phi), F\right) = \mathfrak{T}) 
\end{equation}
\begin{equation} \label{lem3}
 (\mathfrak{v}\left(I(\phi), F\right) = \mathfrak{F}) \Leftrightarrow (\mathfrak{v}\left(I(\phi), T\right) = \mathfrak{T}) 
\end{equation}

Consider the following sentence:

\begin{equation} \label{psi_def}
\psi := (\mathfrak{v}\left(I(\psi), F\right) = \mathfrak{T}) 
\end{equation}

Since $\psi$ is a statement about an equivalence relation, we must also account for a surreptitiously embedded assumption in classical logic, namely, the universality of truth/falsity in characterizing the correctness of any statement, regardless of whether or not it features an equivalence relation. To show this, we shall also assume the following about the statement $\psi$:

\begin{equation} \label{psi_assump1}
 (\mathfrak{v} \left(I(\psi), T\right) = \mathfrak{T})  \Leftrightarrow (\mathfrak{v}\left(I(\psi), F\right) = \mathfrak{T}) 
\end{equation}
\begin{equation} \label{psi_assump2}
(\mathfrak{v}\left(I(\psi), F\right) = \mathfrak{T})  \Leftrightarrow  (\mathfrak{v}\left(I(\psi), F\right) = \mathfrak{F} )
\end{equation}

More specifically, we are assuming that if it is the case that the interpretation of $\psi$ is false, then it satisfies the definition of $\psi$, which means that it is the case that the interpretation of $\psi$ is true, and vice versa. And, if it is not the case that $\psi$ is false, then the definition of $\psi$ is not satisfied (the criterion being the falsity of $\psi$), which means that it is the case that the interpretation of $\psi$ being false is true. We can immediately see that this contradicts the criteria set by the law of the exclusive middle. But let us continue by demonstrating the contradiction explicitly.

\begin{proof}
\begin{align*}
\psi &:= (\mathfrak{v}(I(\psi), F) = \mathfrak{T}) & \text{(Assumption)} \\
(\mathfrak{v}(I(\psi), F) =  \mathfrak{T})
&\Rightarrow (\mathfrak{v}(I(\psi), T) = \mathfrak{T}) & \text{(see \eqref{psi_assump1})} \\
(\mathfrak{v}(I(\psi), F)= \mathfrak{T}) &\Rightarrow (\mathfrak{v}(I(\psi), T) = \mathfrak{F}) & \text{(see \eqref{lem2})} \\
&\bot & \text{(Contradiction)}
\end{align*}
\end{proof}

We have now arrived at a contradiction in the equivalence relation between the interpretation on a sentence letter and its corresponding semantic property. Similarly, let us assume that $\psi$ is interpreted to be false:

\begin{proof}
\begin{align*}
\psi &:= (\mathfrak{v}(I(\psi), F) = \mathfrak{F}) & \text{(Assumption)} \\
(\mathfrak{v}(I(\psi), F) = \mathfrak{F})&\Rightarrow (\mathfrak{v}(I(\psi), F) = \mathfrak{T}) & \text{(see \eqref{psi_assump2})} \\
&\bot & \text{(Contradiction)}
\end{align*}
\end{proof}

Since $\psi$ can only be interpreted as being either true or false, we have exhausted all possible values and have shown that both lead to contradictions. The problem with the law of the excluded middle is that it cannot adequately account for statements that feature self-referential equivalence relations, though we can reconcile with these apparent paradoxes by featuring more nuanced values that reflect other semantic properties which cannot be adequately characterized as being universally true or false but occupy an entirely different epistemological realm.

An ostensible paradox appears to be present in this reasoning. To prove that the law of the excluded middle does not work, we appear to have assumed that it does work for the equivalence valuation function, where an equivalence relation between an interpretation of a sentence and its semantic property either holds or does not hold. However, we have effectively prevented such a paradox by not falling into the fallacy of conflating the codomain of the interpretational function with the codomain of the equivalence valuation function. In other words, $equivalence$ and $truth$ are fundamentally distinct.

\section{Formalization of the Action Axiom}

The introduction of a new truth value in the metalanguage has practical implications. Consider an economic agent who endeavors to raise their utility by selecting an action from the set of all possible actions that can be taken to improve welfare: \(\{A_j\}_{\forall j|j\in N}\). The interpretational utility function, \(I(A_i)\), interprets the action and determines whether it will genuinely improve (i.e. \(I(A_i)=T\)), degenerate (i.e. \(I(A_i)=F\)), or have an indeterminate effect on economic circumstance (i.e. \(I(A_i)=U\)). For the purposes of this demonstration let us assume that the cardinality of the action set is 2: in other words, the agent can choose one of the following: \(\mathbb{A}=\{A_0,A_1\}\).

A set of practical restrictions on behavior is expected; when one action is perceived to be conducive to the outcome of higher utility as opposed to the other action, which is interpreted to be false, then such an action shall be selected. When both actions result in equivalent outcomes, then either action shall be selected, since either consequence is unavoidable, resulting in an ambiguous state. If one action will lead to an unknown outcome whereas the other one will make the agent worse off, then the first action shall be selected. If one action leads to an unknown outcome but the other one will result in a positive outcome, then the agent may or may not select the action that results in the positive outcome depending on one’s preference for risk. The action matrix for a risk-averse agent is shown in the following table:

\begin{center}
Informal Action Selection Matrix
\end{center}
\begin{center}

\begin{tabular}{|c|c|c|c|}
\hline
\(A_1\backslash A_0\) & \(I(A_0)=T\) & \(I(A_0)=F\) & \(I(A_0)=U\) \\
\hline
\(I(A_1)=T\) & \(A_1 \lor A_0\) & \(A_1\) & \(A_1\) \\
\hline
\(I(A_1)=F\) & \(A_0\) & \(A_1 \lor A_0\) & \(A_0\) \\
\hline
\(I(A_1)=U\) & \(A_0\) & \(A_1\) & \(A_1 \lor A_0\) \\
\hline
\end{tabular}
\end{center}

Consider two binary connectives that take the truth values of the actions as inputs and output another truth value to show whether a particular action has been chosen. For example, \(A_1 \oplus_{A_1} A_0\) (i.e. \(\oplus_{A_1} (A_1,A_0)\)) outputs a truth value to note whether action 1 will $necessarily$ be selected by the agent given the underlying values of \(A_1\) and \(A_0\); \(\oplus_{A_0} (A_1,A_0)\) performs the exact same function for action 0. The formal definitions of the connectives are presented in the following tables:
\[
\text{A}_1 \, \oplus_{\text{A}_1} \, \text{A}_0
\]
\[
\begin{array}{|c|c|c|c|}
\hline
A_1\backslash A_0 & I(A_0)=T & I(A_0)=F & I(A_0)=U \\
\hline
I(A_1)=T & V_I (A_1  \oplus_{A_1} A_0 )=u & V_I (A_1  \oplus_{A_1} A_0 )=t & V_I (A_1  \oplus_{A_1} A_0 )=t \\
\hline
I(A_1)=F & V_I (A_1  \oplus_{A_1} A_0 )=f & V_I (A_1  \oplus_{A_1} A_0 )=u & V_I (A_1  \oplus_{A_1} A_0 )=f \\
\hline
I(A_1)=U & V_I (A_1  \oplus_{A_1} A_0 )=f & V_I (A_1  \oplus_{A_1} A_0 )=t & V_I (A_1  \oplus_{A_1} A_0 )=u \\
\hline
\end{array}
\]
\[
\text{A}_1 \oplus_{\text{A}_0} \text{A}_0
\]
\[
\begin{array}{|c|c|c|c|}
\hline
A_1\backslash A_0 & I(A_0)=T & I(A_0)=F & I(A_0)=U \\
\hline
I(A_1)=T & V_I (A_1  \oplus_{A_0} A_0 )=u & V_I (A_1  \oplus_{A_0} A_0 )=f & V_I (A_1  \oplus_{A_0} A_0 )=f \\
\hline
I(A_1)=F & V_I (A_1  \oplus_{A_0} A_0 )=t & V_I (A_1  \oplus_{A_0} A_0 )=u & V_I (A_1  \oplus_{A_0} A_0 )=t \\
\hline
I(A_1)=U & V_I (A_1  \oplus_{A_0} A_0 )=t & V_I (A_1  \oplus_{A_0} A_0 )=f & V_I (A_1  \oplus_{A_0} A_0 )=u \\
\hline
\end{array}
\]

To formalize the contents of this matrix, let us consider two sets that specify the circumstances under which an action is selected. The sets are named \( \Gamma_{A_1} \) and \( \Gamma_{A_0} \) and each logically entails the selection of their respective actions.
\begin{align*}
\Gamma_{A_1} := \{ & \mathfrak{v}(((I(A_1), T) \land ((I(A_0), F) \lor (I(A_0), U))) \\
& \lor ((I(A_1), U) \land (I(A_0), F))) = \mathfrak{T} \}
\end{align*}
\begin{align*}
\Gamma_{A_0} := \{ & \mathfrak{v}(((I(A_0), T) \land ((I(A_1), F) \lor (I(A_1), U))) \\
& \lor ((I(A_0), U) \land (I(A_1), F))) = \mathfrak{T} \}
\end{align*}\\
\textbf{Annotated Proof of $\Gamma_{A_1} \models \mathfrak{v}(V_I (A_1 \oplus_{A_1} A_0), t)=\mathfrak{T}$:}
\begin{proof}
\begin{align*}
\overbrace{P_1}^{\text{Premise 1}} & := \underbrace{\mathfrak{v}(((I(A_1), T) \land ((I(A_0), F) \lor (I(A_0), U)))}_{\text{Condition A}} \\
& \quad \lor \underbrace{((I(A_1), U) \land (I(A_0), F))) = \mathfrak{T},}_{\text{Condition B}} \\
\overbrace{P_2}^{\text{Premise 2}} & := \underbrace{\mathfrak{v}(V_I (A_1 \oplus_{A_1} A_0), t)=\mathfrak{F},}_{\text{Assertion}} \\
P_1 & \Rightarrow \overbrace{(\mathfrak{v}((I(A_1) ,T) \land ((I(A_0), F) \lor (I(A_0), U))) = \mathfrak{T})}^{\text{Consequence 1A}} \\
& \quad \lor \underbrace{(\mathfrak{v}((I(A_1), U) \land (I(A_0), F)) = \mathfrak{T}),}_{\text{Consequence 1B}} \\
P_1 & \Rightarrow \overbrace{((\mathfrak{v}((I(A_1), T) \land (I(A_0), F)) = \mathfrak{T})}^{\text{Consequence 1.1A}} \\
& \quad \lor \underbrace{(\mathfrak{v}((I(A_1), T) \land (I(A_0), U)) = \mathfrak{T}))}_{\text{Consequence 1.2A}} \\
& \quad \lor \underbrace{(\mathfrak{v}((I(A_1), U) \land (I(A_0), F)) = \mathfrak{T}),}_{\text{Consequence 1B}} \\
P_1 & \Rightarrow \overbrace{(\mathfrak{v}((V_I (A_1 \oplus_{A_1} A_0),t) \lor (V_I (A_1 \oplus_{A_1} A_0), t))=\mathfrak{T})}^{\text{Final Conclusion via Discharge of A}} \\
& \quad \lor \underbrace{(\mathfrak{v}(V_I (A_1 \oplus_{A_1} A_0), t)=\mathfrak{T})}_{\text{Final Conclusion via Discharge of B}} \\
P_1 & \Rightarrow \overbrace{\mathfrak{v}(V_I (A_1 \oplus_{A_1} A_0), t)=\mathfrak{T}}^{\text{Final Conclusion}} \\
P_1 & \Rightarrow \underbrace{\lnot P_2.}_{\text{Contradiction}}
\end{align*}
Therefore, $\Gamma_{A_1} \models \mathfrak{v}(V_I (A_1 \oplus_{A_1} A_0), t)=\mathfrak{T}$.
\end{proof}
\textbf{Proof of $\Gamma_{A_0} \models \mathfrak{v}(V_I (A_1 \oplus_{A_0} A_0), t)=\mathfrak{T}$:}
\begin{proof}
\begin{align*}
P_3 & := \mathfrak{v}(((I(A_0), T) \land ((I(A_1), F) \lor (I(A_1), U))) \\
& \lor ((I(A_0), U) \land (I(A_1), F))) = \mathfrak{T}, \\
P_4 & := \mathfrak{v}(V_I (A_1 \oplus_{A_0} A_0), t)=\mathfrak{F}, \\
P_3 & \Rightarrow (\mathfrak{v}((I(A_0), T) \land ((I(A_1), F) \lor (I(A_1), U))) = \mathfrak{T}) \\
& \lor (\mathfrak{v}((I(A_0), U) \land (I(A_1), F)) = \mathfrak{T}), \\
P_3 & \Rightarrow ((\mathfrak{v}((I(A_0), T) \land (I(A_1), F)) = \mathfrak{T}) \\
& \lor (\mathfrak{v}((I(A_0), T) \land (I(A_1), U)) = \mathfrak{T})) \\
& \lor (\mathfrak{v}((I(A_0), U) \land (I(A_1), F)) = \mathfrak{T}), \\
P_3 & \Rightarrow (\mathfrak{v}((V_I (A_1 \oplus_{A_0} A_0), t) \lor (V_I (A_1 \oplus_{A_0} A_0), t))=\mathfrak{T}) \\
& \lor (\mathfrak{v}(V_I (A_1 \oplus_{A_0} A_0), t)=\mathfrak{T}), \\
P_3 & \Rightarrow\mathfrak{v}(V_I (A_1 \oplus_{A_0} A_0), t)=\mathfrak{T}, \\
P_3 & \Rightarrow \lnot P_4.
\end{align*}
Therefore, $\Gamma_{A_0} \models \mathfrak{v}(V_I (A_1 \oplus_{A_0} A_0), t)=\mathfrak{T}$.
\end{proof}

Hence, the selection criterion is as follows: if there exists a $\Gamma$ such that for all $ \mathfrak{v} \in \Gamma$, $ \mathfrak{v} = \mathfrak{T}$, then $V_I (A_1 \oplus_a A_0) = t$. This is an important assumption. It states that the selection of human actions is based on whether they are conducive to the realization of desired outcomes. We may now expand upon this analysis by showing that an agent may respond to an objective reality based on one’s subjective perception; this, in turn, is shaped by the objective consequences of the actions selected.

\subsection{Intuition behind Proofs}

As the proofs show, the equivalence valuation function plays a substantial role in making deductive inferences. Essentially, the preceding proofs start from assertions regarding what the truth values of the atomic sentences are, and proceeds to show that the featured combinations of the atomic truth values, when considered alongside the valuation on a particular connective, would result in the selection of a particular action. 

\section{Dynamic Six-Valued Logic System }
\subsection{Motivation and Basic Ideas}
Having established a system in which an agent may weigh the relative merits of each action before making a unique selection, we may now turn to a system in which we distinguish between subjective perceptions and objective realities. Subjective perceptions play a significant role in economic theory. The subjective theory of value, for instance, posits that the value of any good is determined by the eye of the beholder, or the perceived utility that can be derived from it (this is a tautology; utility is perceived as it is a subjective experience). As opposed to saying that an agent makes a decision on the basis of objective circumstances, it is more accurate for us to depict the agent's decisions as a consequence of his or her perception of these circumstances. Once an action has been selected, both the agent's belief regarding that action, and the objective reality of whether an action is conducive to a particular objective, shall be updated, and the selection process iterates (hence giving rise to a $dynamic$ system). 

\subsection{Introduction}

Following from our previous assumptions regarding the selection of a particular course of action: consider a set, $\mathbb{T}$, of six truth values, namely, subjectively true ($T_s$), subjectively false ($F_s$), subjectively unknown ($U_s$), objectively true ($T_o$), objectively false ($F_o$), and objectively unknowable ($U_o$):
\begin{center}
$\mathbb{T}=\{T_s,F_s,U_s,T_o,F_o,U_o\}$
\end{center}

It is assumed that the agent now selects one action from the action set, which contains at least two possible actions. In other words: 

\[
\mathbb{A} = \{A_j | j \in \mathbb{N}, j \geq 2\}
\]

An interpretation of an action is a function that maps an element of the set $A$ to two truth values, each taken from a partition of $\mathbb{T}$.
\begin{center}
$S=\{T_s,F_s,U_s \}$

$O=\{T_o,F_o,U_o\}$

$\forall i, \forall t,\exists! s \in S, \exists! o \in O: \mathfrak{v}(I_t (A_i),[s/o])=\mathfrak{T}$
\end{center}

A set of j-nary connectives, $\oplus$, is defined. Each connective determines whether an action is selected on the basis of the interpretational value of all possible actions.
\[
\oplus=\left\{ \begin{array}{@{}l@{}}
\oplus_{A_0} (A_0, A_1, \dots, A_j)\\
\quad \quad \quad \quad \vdots \\
\oplus_{A_j} (A_0, A_1, \dots, A_j)
\end{array} \right\}
\]
The codomain of the valuation of each connective is $O$. This is because the selection of an action can only be interpreted as an objective realization of individual behavior. 
\[
\forall I_t, \forall \oplus_{(A\in \mathbb{A})}, V_{I_t} (\oplus_{(A\in \mathbb{A})} (A_0, A_1, \ldots, A_j)) \in O
\]
For each individual connective, 
\begin{align*}
& \mathfrak{v}(V_{I_t} (\oplus_{(A_m\in \mathbb{A})} (A_0, A_1, \ldots, A_j)), T_o)=\mathfrak{T} \Leftrightarrow \exists! A_m\in \mathbb{A} : \forall s\in\{F_s, U_s\}, \forall o_0, o_1\in O, \\
& \forall A_n\in \mathbb{A}: A_n \neq A_m, ((\mathfrak{v}(I_t (A_m ),[T_s/o_0])=\mathfrak{T})\land(\mathfrak{v}(I_t (A_n ),[s/o_1])=\mathfrak{T})) \\
& \quad \lor ((\mathfrak{v}(I_t (A_m ),[U_s/o_0])=\mathfrak{T})\land(\mathfrak{v}(I_t (A_n ),[F_s/o_1])=\mathfrak{T}))
\end{align*}

More succinctly, an action is selected if its ability to contribute to welfare is subjectively true whereas that of all other alternatives is subjectively false or unknown, or if its ability is subjectively unknown whereas that of all other alternatives is subjectively false. 
\subsection{Belief Update Mechanism}
We have established that the selection of an action is solely predicated on its subjective interpretation. However, a subjective truth value can be morphed by an objective outcome, which is inhered in the objective truth value. Additionally, the objective truth values of all actions can be morphed through the selection of an action. This gives rise to a dynamic logical system where beliefs can be updated by the feedback of an environment. 
\begin{align*}
    & \forall t, \forall a, b, c, d, e, f \in \{T, F, U\}, \forall A_n, A_m \in \mathbb{A} : A_n \neq A_m, \\
    & (\mathfrak{v}(V_{I_t} (\oplus_{A_m} (A_0, A_1, \ldots, A_j)), T_o)=\mathfrak{T}) \Rightarrow \\
    & (((\mathfrak{v}(I_t (A_n), [a_s / b_o]) = \mathfrak{T}) \Rightarrow (\mathfrak{v}(I_{(t+1)} (A_n), [a_s / c_o])=\mathfrak{T})) \land \\
    & \quad ((\mathfrak{v}(I_t (A_m), [d_s / e_o]) = \mathfrak{T}) \Rightarrow (\mathfrak{v}(I_{(t+1)} (A_m), [e_s / f_o])=\mathfrak{T})))
\end{align*}
Note that this definition relies on a strict axiom of preservation. In other words, it relies on the fact that beliefs do not spontaneously change and are preserved unless altered through the manifestation of an action. In contrast, an omniscient agent is privy to the objective truth values of all actions at any point in time and does not need to allow the consequence of the action to manifest to adjust their beliefs. In other words: 
\begin{align*}
    & \forall t, \forall a, b, c, d, e, f \in \{T, F, U\}, \forall A_n, A_m \in \mathbb{A} : A_n \neq A_m, \\
    & (\mathfrak{v}(V_{I_t} (\oplus_{A_m} (A_0, A_1, \ldots, A_j)), T_o)=\mathfrak{T}) \Rightarrow \\
    & (((\mathfrak{v}(I_t (A_n), [a_s / a_o]) = \mathfrak{T}) \Rightarrow (\mathfrak{v}(I_{(t+1)} (A_n), [b_s / b_o])=\mathfrak{T})) \land \\
    & \quad ((\mathfrak{v}(I_t (A_m), [c_s / c_o]) = \mathfrak{T}) \Rightarrow (\mathfrak{v}(I_{(t+1)} (A_m), [d_s / d_o])=\mathfrak{T})))
\end{align*}
However, this is an eminently unrealistic assertion in practice. A realistic model should incorporate latency and discrepancies to arrive at a more accurate depiction of human action. 
Returning to the preservative model, the previous selection process continues unless the following occurs: 
\begin{align*}
& \forall k \in \mathbb{N}, \mathfrak{v}(V_{I_{(t+k)}} (\oplus_{(A_m\in \mathbb{A})} (A_0, A_1, \ldots, A_j)), T_o)=\mathfrak{T} \Leftrightarrow \exists! A_m\in \mathbb{A} : \forall k \in \mathbb{N}, \forall s\in\{F_s, U_s\}, \\
& \forall o_0, o_1\in O, \forall A_n\in \mathbb{A}: A_n \neq A_m, ((\mathfrak{v}(I_{(t+k)} (A_m ),[T_s/o_0])=\mathfrak{T})\land(\mathfrak{v}(I_{(t+k)} (A_n ),[s/o_1])=\mathfrak{T})) \\
& \quad \lor ((\mathfrak{v}(I_{(t+k)} (A_m ),[U_s/o_0])=\mathfrak{T})\land(\mathfrak{v}(I_{(t+k)} (A_n ),[F_s/o_1])=\mathfrak{T}))
\end{align*}

This can be characterized as an equilibrium state in which the further selection of $A_m$  will not result in a different objective truth value: resulting in it being selected indefinitely and irrespective of the objective truth values of all other actions.
\subsection{Discussion}
So far, we have neglected cases where the agent is confronted with at least two actions that are both perceived to be conducive to a particular objective or when the subjective truth values for all actions are false or unknown. This is because in such cases there does not exist a rational basis for selection. Contemporary economic methods purport to address this by assigning weightings to actions, thereby conveniently denoting the degree to which each action is conducive to particular outcomes. But even these methods suggest that the agent should be indifferent between virtually identical alternatives that render similar levels of utility. To address this, an indifference curve analysis can be employed to discriminate between such outcomes. If each action incurs a cost, the outcome that yields the maximal level of utility given a particular budget is chosen. It is possible for us to formalize this notion in the metalanguage by assigning truth values on a nuanced criterion. For example, a sentence letter, $A_i$, can be interpreted as asserting: ``Given a budget, $B$, allocating $X$ dollars to purchase good A, where $X \leq B$, and $(B-X)$ dollars to purchase good B would allow us to reach the maximal amount of utility derivable from B." Whereas $A_{i+1}$ can be interpreted as saying: ``Given a budget, $B$, allocating $(X+1)$ dollars to purchase good A, where $(X+1) \leq B$, and $(B-(X+1))$ dollars to purchase good B would allow us to reach the maximal amount of utility derivable from B." In some instances, this may mathematically preclude the truth value $U$, since the indifference curve approach is inherently deterministic and offers no room for absolute ambiguity. Yet again, this is not necessarily true. For example, if the marginal rate of substitution were increasing and the indifference curve function which renders the maximal level of possible utility derivable from the budget is concave, it would be possible for a linear budget constraint function to reach the concave curve at more than one point, resulting in an ambiguous state for which no unique allocation of one budget is present. 
\subsubsection{Example}
As a demonstration, consider an example of an agent who is interested in maximizing their utility through the selection of two perfectly complementary goods. There are two actions in our model: either the agent can select good A (\( A_0 \)), or it can select good B (\( A_1 \)). Let us suppose that the objective truth values for conducting either action are initially false. This is because selecting either good on its own would not allow the agent to realize the utility of their purchase.

The agent is initially skeptical of the need to purchase a complementary good. In other words: 
\[
\mathbb{A}=\{A_0,A_1\}
\]
\[
\mathfrak{v} (I_0 (A_0 ),[U_s/F_o])=\mathfrak{T}
\]
\[
\mathfrak{v}(I_0 (A_1 ),[T_s/F_o])=\mathfrak{T}
\]

We now consider two connectives that specify whether or not a particular action is selected, namely, \( \oplus_{A_0} \) and \( \oplus_{A_1} \). We have previously specified that the valuation on such a connective is objectively true under the following circumstances:

{\small\noindent \textbf{Recall} from 6.2:}
\begin{align*}
& \mathfrak{v}(V_{I_t} (\oplus_{(A_m\in \mathbb{A})} (A_0, A_1)), T_o)=\mathfrak{T} \Leftrightarrow \exists! A_m\in \mathbb{A} : \forall s\in\{F_s, U_s\}, \forall o_0, o_1\in O, \\
& \forall A_n\in \mathbb{A}: A_n \neq A_m, ((\mathfrak{v}(I_t (A_m ),[T_s/o_0])=\mathfrak{T})\land(\mathfrak{v}(I_t (A_n ),[s/o_1])=\mathfrak{T})) \\
& \quad \lor ((\mathfrak{v}(I_t (A_m ),[U_s/o_0])=\mathfrak{T})\land(\mathfrak{v}(I_t (A_n ),[F_s/o_1])=\mathfrak{T}))
\end{align*}

Since the interpretation of \( A_0 \) when \( t=0 \) is \( [U_s / F_o] \) and that of \( A_1 \) is \( [T_s / F_o] \), \( V_I (\oplus_{A_1} (A_0,A_1))= T_o \), hence, \( A_1 \) is selected. Upon selecting this action, the agent’s subjective belief collapses with the objective truth value of the action: 

{\small\noindent \textbf{Recall} from 6.3:}
\begin{align*}
    & \forall t, \forall a, b, c, d, e, f \in \{T, F, U\}, \forall A_n, A_m \in \mathbb{A} : A_n \neq A_m, \\
    & (\mathfrak{v}(V_{I_t} (\oplus_{A_m} (A_0, A_1)), T_o)=\mathfrak{T}) \Rightarrow \\
    & (((\mathfrak{v}(I_t (A_n), [a_s / b_o]) = \mathfrak{T}) \Rightarrow (\mathfrak{v}(I_{(t+1)} (A_n), [a_s / c_o])=\mathfrak{T})) \land \\
    & \quad ((\mathfrak{v}(I_t (A_m), [d_s / e_o]) = \mathfrak{T}) \Rightarrow (\mathfrak{v}(I_{(t+1)} (A_m), [e_s / f_o])=\mathfrak{T})))
\end{align*}

Since \( A_1 \) is selected, let \( I_1 (A_1 )=[F_s / F_o] \). In other words, upon selecting the action, its consequences manifest, and the agent realizes that the purchase of good B is insufficient. On the other hand, let \( I_1 (A_0 )=[U_s / T_o] \). This implies that the selection of \( A_1 \) has affected the objective truth value of \( A_0 \). Since the subjective truth value of \( A_0 \) is unknown, whereas that of \( A_1 \) is false, \( A_0 \) will be selected. This results in \( I_2 (A_0 )=[T_s / T_o] \). The process of selection now comes to a meaningful halt because the subjective truth value of the selected action now reflects its objective truth value, which is true (i.e. conducive to the objective of realizing utility).

\section{Theories of Choice and Alternative Forms of Consequence Realization}

We may now consider the diverse ways in which we could define the behavior of \( I_t \), and correspondingly, \( V_{I_t} \), across \( t \). We have previously assumed that upon selecting an arbitrary action, \( A_m \), the subjective truth values of all other actions, \( a_s \), are preserved at \( I_{t+1} \). Additionally, we have also assumed that the only way in which an action can manifest its objective truth value is by selecting that action. However, it is also possible for the subjective truth values of other un-selected actions to change on the basis of the realization of an objective truth value of any particular action, whether it has been selected or not. In other words, it could be useful to consider a lenient form of consequence realization:
\[
\forall t, \forall a,b,c,d\in\{T,F,U\},  \forall A_n\in \mathbb{A}, (\mathfrak{v}(I_t (A_n ),[a_s/b_o])=\mathfrak{T}) \Rightarrow (\mathfrak{v}(I_{t+1} (A_n ),[c_s/d_o])=\mathfrak{T})
\]

On its own, this definition does not offer any meaningful information on how we can make predictions regarding dynamic beliefs or changing realities. Hence, for us to continue our analysis it becomes necessary for us to formalize the circumstances in which an action is selected and the corresponding architecture of the individual’s belief. 

\subsection{Monty Hall Problem}
An example that illustrates the need to formalize the “theories” under which actions are selected is the Monty Hall Problem. Consider being in a game show where three doors are present before you; a goat lies behind two doors whereas a car lies behind the third. The objective of the utility-maximizing agent is to select a door that results in the obtainment of the car. Upon selecting a door, Monty Hall reveals one of the doors that did not have a car behind it. You are then given the choice of either selecting your current door or selecting the other door which has not yet been revealed or selected.

Let us define each action in the following manner: 
\( \mathbb{A}=\{D_1,D_2,D_3\} \)

Where \( D_i \) is syntactically equivalent to saying that the act of selecting the \( i^{th} \) door would result in the obtainment of the car. Let us suppose that the interpretational truth value of each action is informally defined: 
\[
\begin{aligned}
& \mathfrak{v}(I_0 (D_1 ),[U_s⁄F_s /T_o ])=\mathfrak{T} \\
& \mathfrak{v}(I_0 (D_2 ),[U_s⁄F_s /F_o ])=\mathfrak{T} \\
& \mathfrak{v}(I_0 (D_3 ),[T_s/F_o])=\mathfrak{T}
\end{aligned}
\]

It is assumed that the action we have selected, namely, \( D_3 \), is subjectively true (whereas the subjective truth values of \( D_1 \) and \( D_2 \) are unknown, or false) precisely because the game requires us to select an action over others, thereby revealing our preference. However, upon selecting the action, the objective truth value of selecting the 3rd door does not manifest itself directly. On the contrary, we become privy to the objective truth value of the second door, which is revealed to be false. The updated truth values are as follows. 
\[
\begin{aligned}
& \mathfrak{v}(I_1 (D_1 ),[U_s⁄F_s /T_o ])=\mathfrak{T} \\
& \mathfrak{v}(I_1 (D_2 ),[F_s/F_o])=\mathfrak{T} \\
&  \mathfrak{v}(I_1 (D_3 ),[T_s/F_o])=\mathfrak{T}
\end{aligned}
\]

According to our previous set of axioms, we have demonstrated that when an agent believes that an action leads to a particular outcome, then the agent should have no reason to spontaneously alter their beliefs unless the consequences of that action materialize and are revealed to be contrary to what is expected. In other words, if the agent’s conviction in their initial selection (3rd door) is genuinely true, the agent should have no reason to then select the first door: which they initially did not select because its subjective truth value was false/unknown as opposed to being true. This runs athwart the conventional result from probabilistic reasoning, namely: if the likelihood of initially selecting a goat is comparatively high, then the likelihood of being able to obtain a car upon the reversal of one’s initial selection would be comparatively high as well. Thus, from an epistemological perspective, the selection of an action is not necessarily a reflection of one’s faith in its ability to be conducive to a particular outcome. Moreover, upon selecting an action, it may not manifest its objective truth value directly but result in the actualization of the objective truth values of other courses of action, which in turn affect the subjective truth values of the remaining set of actions (despite the fact that they have not been selected). 
\subsection{Overview of Preceding Theory}
We have previously stated that when an agent is confronted with at least two actions that are both perceived to be conducive to a particular objective or when the subjective truth values of all actions are false or unknown, there does not exist a rational basis for selection. The definition of rationality espoused is limited to cases where we can denote an agent’s clear preference for one course of action over another. To gain a comprehensive view of the theory we have developed thus far, consider a set, \( \Gamma_R \), which denotes the theory for strictly rational behavior under belief preservation: 
\[
\begin{aligned}
\Gamma_R := \Bigg\{ & 
\begin{aligned}[t]
& \overbrace{{\mathbb{A} = \{{A_j | j \in \mathbb{N}, j \geq 2\}}}}^{\text{Action Set}}, 
\overbrace{{S=\{{T_s,F_s,U_s \}}}}^{\text{Subjective Faith}}, 
\overbrace{{O=\{{T_o,F_o,U_o\}}}}^{\text{Objective Realities}}, 
\\
& \underbrace{{(\forall i, \forall t,\exists! s \in S, \exists! o \in O: \mathfrak{v}(I_t (A_i),[s/o])=\mathfrak{T})}}_{\text{Interpretation of an action}}, 
\\
& \overbrace{{\oplus=\left\{ \begin{array}{@{}l@{}}
\oplus_{A_0 } (A_0,A_1,\ldots,A_j)\\
\quad \quad \quad \quad \vdots \\
\oplus_{A_j } (A_0,A_1,\ldots,A_j)
\end{array} \right\}}}^{\text{Action Selector}} 
\end{aligned} \\
& \begin{aligned}[t]
& , \underbrace{{(\forall I_t,\forall \oplus_{(A\in \mathbb{A})},V_{I_t} (\oplus_{(A\in \mathbb{A})} (A_0,A_1,\ldots,A_j)) \in O)}}_{\text{Selections have an Objective Presence}}, 
\\
& \overbrace{(\mathfrak{v}(V_{I_t} (\oplus_{(A_m\in \mathbb{A})} (A_0, A_1, \dots, A_j)), T_o)=\mathfrak{T})}^{\text{Outcome: Selection of $A_m$}}  \Leftrightarrow (\exists!A_m\in \mathbb{A}:\forall s\in\{F_s,U_s \},\forall o_0,o_1\in O,\\
&\underbrace{\forall A_n\in \mathbb{A}:A_n\neq A_m}_{\text{Specifies non-selected action}},\underbrace{(({\mathfrak{v}(I_t (A_m ),[T_s/o_0])=\mathfrak{T}})\land(\mathfrak{v}(I_t (A_n ),[s/o_1])=\mathfrak{T}))}_{\text{First condition for selection}}\lor \\
& \overbrace{((\mathfrak{v}(I_t (A_m ),[U_s/o_0])=\mathfrak{T})\land(\mathfrak{v}(I_t (A_n ),[F_s/o_1])=\mathfrak{T}))}^{\text{Alternative condition for selection}})
\end{aligned} \\
& \begin{aligned}[t]
& (\forall t, \forall a,b,c,d,e,f\in\{T,F, U\} , \forall A_n, A_m\in \mathbb{A}:A_n\neq A_m,\\
& (\overbrace{{\mathfrak{v}(V_{I_t} (\oplus_{(A_m\in \mathbb{A})} (A_0, A_1, \dots, A_j)), T_o)=\mathfrak{T}}}^{\text{Condition: If an action is selected...}}) \Rightarrow (((\overbrace{{\mathfrak{v}(I_t (A_n ),[a_s/b_o])=\mathfrak{T})}}^{\text{Specifies initial value}} \Rightarrow \\
& \underbrace{(\mathfrak{v}(I_{(t+1)} (A_n ),[a_s/c_o])=\mathfrak{T}))}_{\text{Specifies updated value}} \land \underbrace{((\mathfrak{v}(I_t (A_m ),[d_s/e_o])=\mathfrak{T})\Rightarrow (\mathfrak{v}(I_{(t+1)} (A_m),[e_s/f_o])=\mathfrak{T}))}_{\text{Transmission of $e_o$ to $e_s$ via selection}}))\Bigg\} 
\end{aligned}
\end{aligned}
\]

In concise terms, we have defined the set of actions, the set of possible truth values, the relationship between an interpretation of an action and its corresponding truth value, the semantic properties of a \( j \)-nary connective, and the process by which the selection of an action (i.e. a valuation on an action that results in an objective truth, \( T_o \)) generates updated interpretations.  

The theory for strict rational choice, \( \Gamma_R \), does not address the selection of an action in the following cases, or addresses them trivially by assigning an unknown valuation: 
\[
\begin{aligned}
& \exists t: \forall o_1\in O, \forall A_w\in \mathbb{A},\mathfrak{v}(I_t (A_w),[T_s/o_1])=\mathfrak{T} \\
& \exists t:\forall o_1\in O, \forall A_w\in \mathbb{A}, \mathfrak{v}(I_t (A_w),[F_s/o_1])=\mathfrak{T} \\
& \exists t:\forall o_1\in O, \forall A_w\in \mathbb{A},\mathfrak{v}(I_t (A_w),[U_s/o_1])=\mathfrak{T} \\
& \exists t: \forall j>2, \forall a\in\{F_s,U_s \},  \forall o_1,o_2\in O, \forall A_w,A_n\in \mathbb{A}, \\ 
& \quad \exists d \in \mathbb{N}, j > d \geq 2 : \{A_n\}_{n=1}^{d}
\subset \mathbb{A}:\forall n,A_n\neq A_w, \\
& \quad (\mathfrak{v}(I_t (A_n),[T_s/o_1])=\mathfrak{T})\land(\mathfrak{v}(I_t (A_w),[a/o_2])=\mathfrak{T}) \\
&\exists t: \forall j>2, \forall o_1,o_2\in O, \forall A_w, A_n\in \mathbb{A}, \\
& \quad \exists d \in \mathbb{N}, j > d \geq 2 : \{A_n\}_{n=1}^{d}
\subset \mathbb{A}:\forall n,A_n\neq A_w, \\
& \quad (\mathfrak{v}(I_t (A_n),[U_s/o_1])=\mathfrak{T})\land(\mathfrak{v}(I_t (A_w),[F_s/o_2])=\mathfrak{T})
\end{aligned}
\]

\subsubsection{Translations from formal language}
The following cases are not addressed due to the implicit notion of revealed preference: \\
1. Cases where every action is interpreted as being conducive to a particular objective. \\
2. Cases where every action is not interpreted as being conducive to a particular objective. \\
3. Cases where the interpretation of every action is unknown. \\
4. Cases where the interpretation of at least two actions are true, whereas that of all other actions are false or unknown. \\
5. Cases where the interpretation of at least two actions are unknown, and the rest are interpreted as being false. \\

\textbf{Recall} from section 5 where the definition of a $j$-nary connective is presented, we claimed that such a connective can only show whether it is $necessarily$ the case that a unique action has been selected. If more than one action is conducive to the realization of a particular objective, the agent should have no reason to make a selection on the basis of one's preference, since the $explicit$ $selection$ of an action necessarily implies that the agent $explicitly$ $prefers$ one action over the other. As we have stated in section 6.4, these are cases where there does not exist a strictly rational basis of selection, thereby necessitating alternative theories.

\subsection{Alternative Theory}
The primary justification for the selection of any action is to check whether its objective truth value would prove to be ephemerally true, or robustly and consistently true. However, different philosophical beliefs underpin such methodologies. To demonstrate this, consider how the selection of an action occurs under the belief of philosophical skepticism, that is, the belief that objective truths are unknowable (Klein). Insofar as we are only capable of perceiving the extent to which an action is conducive to a particular outcome through subjective truth values, any method of selection on the basis of this perception must be unreliable. Although this position results in the invalidation of all selection methods, we may consider a similar though less stringent position of empirical skepticism ($\Gamma_{ES}$). Under $\Gamma_{ES}$, an action is selected when its probability of realizing a particular objective exceeds that of all other actions. 
\[
\begin{aligned}
\Gamma_{ES} := \Bigg\{ & 
\begin{aligned}[t]
& \overbrace{{\mathbb{A} = \{{A_j | j \in \mathbb{N}, j \geq 2\}}}}^{\text{Action Set}}, \overbrace{{S=\{{x \in \mathbb{R}|0 \leq x \leq 1  \}}}}^{\text{Subjective Faith}}, \overbrace{{O=\{{y \in \mathbb{R}|0 \leq y \leq 1 \}}}}^{\text{Objective Realities}}, \\
\\
& \underbrace{{(\forall i, \forall t,\exists! s \in S, \exists! o \in O: \mathfrak{v}(I_t (A_i),[s/o])=\mathfrak{T})}}_{\text{Interpretation of an action}}, \\
& \overbrace{{\oplus=\left\{ \begin{array}{@{}l@{}}
\oplus_{A_0 } (A_0,A_1,\ldots,A_j)\\
\quad \quad \quad \quad \vdots \\
\oplus_{A_j } (A_0,A_1,\ldots,A_j)
\end{array} \right\}}}^{\text{Action Selector}}
\end{aligned} \\
& \begin{aligned}[t]
&, \underbrace{{(\forall I_t,\forall \oplus_{(A\in \mathbb{A})},V_{I_t} (\oplus_{(A\in \mathbb{A})} (A_0,A_1,\ldots,A_j)) \in \{T_o,F_o,U_o\} )}}_{\text{Selections have an Objective Presence}}, \\
& \overbrace{{(\mathfrak{v}(V_{I_t} (\oplus_{(A_m\in \mathbb{A})} (A_0, A_1, \dots, A_j)), T_o)=\mathfrak{T})}}^{\text{Outcome: Selection of $A_m$}}  \Leftrightarrow (\exists! A_m\in \mathbb{A}:\forall s_0,s_1\in S , \forall o_0,o_1\in O,\\
&\underbrace{{\forall A_n\in \mathbb{A}:A_n\neq A_m}}_{\text{Specifies non-selected action}},\underbrace{{((\mathfrak{v}(I_t (A_m ),[s_0/o_0])=\mathfrak{T})\land(\mathfrak{v}(I_t (A_n ),[s_1/o_1])=\mathfrak{T}))}}_{\text{Specifies the initial values of a selected and a non-selected action}} \\
&\land \overbrace{{(s_1 > s_0)}}^{\text{Condition}}),(\forall t, \forall a,d,e\in S ,  \forall b,c,e,f\in O, \forall A_n,A_m\in \mathbb{A}:A_n\neq A_m,\\
& (\overbrace{{\mathfrak{v}(V_{I_t} (\oplus_{(A_m\in \mathbb{A})} (A_0, A_1, \dots, A_j)), T_o)=\mathfrak{T}}}^{\text{Condition: If an action is selected...}}) \Rightarrow ((\overbrace{{(\mathfrak{v}(I_t (A_n ),[a/b])=\mathfrak{T})}}^{\text{Specifies initial value}} \Rightarrow \\
& \underbrace{(\mathfrak{v}(I_{(t+1)} (A_n ),[a/c])=\mathfrak{T}))}_{\text{Specifies updated value}} \land \underbrace{((\mathfrak{v}(I_t (A_m ),[d/e])=\mathfrak{T})\Rightarrow (\mathfrak{v}(I_{(t+1)} (A_m),[e/f])=\mathfrak{T}))}_{\text{Transmission of $e$ via selection}}))\Bigg\} 
\end{aligned}
\end{aligned}
\]

\section{Areas of Further Discussion}
One area of further discussion regarding the preceding model is whether government intervention could lead to permanent distortions in a rational agent's subjective perception of his or her objective circumstances, thereby producing inefficient social outcomes. This can be modelled by limiting the type of actions that are featured in the agent's action set, or by inflating the agent's subjective perception of certain actions to artificially high values (to indicate their perceived effectiveness, or in discrete cases, $True$). Admittedly, such a system would go beyond what this paper could possibly cover, and the notations used to express $\Gamma_R$ and $\Gamma_{ES}$ would probably pale in complexity. 

\section{Conclusion}
The current analysis of dynamic many-valued logic is far from complete and deserves further analyses. A variety of applications are predicated on this foundational approach to economic reasoning. Additionally, further analyses could be made by examining an efficient method of defining an updated interpretation on the basis of a selected action, or a series of selected actions; this, however, ultimately rests on the structure of the dynamic problem being investigated.  It is also worth considering a set of sets of actions that exist for each moment of interpretation, $t$, thereby allowing us to model certain decisions that expand or constrict the domain of possible options. The explicit enunciation of the underlying philosophy behind the selection of an action in cases where no action results in a strict preference would allow for the rigorous identification of underlying assumptions, resulting in a robust method of selection. 

\section*{Data Availability}
This work does not generate any data sets because it proceeds within a theoretical and mathematical approach. One can obtain the relevant materials from the references below. 

\section*{References}
1. Falkenstein, Lorne, et al. Logic Works : A Rigorous Introduction to Formal Logic. New York, Routledge, Taylor and Francis Group, 2022.\\
2. Klein, Peter. “Skepticism (Stanford Encyclopedia of Philosophy).” Stanford.edu, 2015, plato.stanford.edu/entries/skepticism/.
\end{document}